%% file: main_file.tex
\documentclass[3p,times,procedia]{elsarticle}
\usepackage{ecrc}
\usepackage[bookmarks=false]{hyperref}
    \hypersetup{colorlinks,
      linkcolor=blue,
      citecolor=blue,
      urlcolor=blue}

\volume{00}

\firstpage{1}

\journalname{Procedia Computer Science}

\runauth{Author name}

\jid{procs}

\usepackage{amssymb}
\usepackage{tabularx}

\usepackage[figuresright]{rotating}

\begin{document}
\begin{frontmatter}



\dochead{The 17th International Conference on Ambient Systems, Networks and Technologies (ANT) \\ April 14-16, 2026, Istanbul, Türkiye}%

\title{A Lay User Explainable Food Recommendation System Based on Hybrid Feature Importance Extraction and Large Language Models}


\author[e]{Melissa Tessa}
\author[a]{Diderot D. Cidjeu \corref{cor1}} 
\author[b]{Rachele Carli}
\author[e]{Sarah Abchiche}
\author[d]{Ahmad Aldarwish}
\author[c]{Igor Tchappi}

\author[f]{Amro Najjar }

\address[a]{Centre de Recherche Panafricain en Management pour le Développement(CERPAMAD), 06 BP 9258, Ouagadougou, Burkina Faso}
\address[b]{Responsible AI group, Umeå University, SE-901 87 Umeå, Sweden }
\address[c]{FINATRAX, SnT, University of Luxembourg, 2 Av. de l'Universite L-4365, Esch-sur-Alzette, Luxembourg}
\address[d]{Devoteam, 7101 Prince Abdulaziz Ibn Musaid Ibn Jalawi St, As Sulimaniyah, DISTRICT 12232, Riyadh, Kingdom of Saudi Arabia}
\address[e]{Ecole Nationale Supérieure d’Informatique d’Alger, ESI ex-INI Alger, Algeria}
\address[f]{Luxembourg Institute of Science and Technology, Maison de l'innovation 5, avenue des Hauts-Fourneaux, L-4362 Luxembourg}

\begin{abstract}
Large Language Models (LLM) have experienced strong development in recent years, with varied applications. This paper uses LLMs to develop a post-hoc process that provides more elaborated explanations of the results of food recommendation systems. By combining LLM with a hybrid extraction of key variables using SHAP, we obtain dynamic, convincing and more comprehensive explanations to lay user, compared to those in the literature. This approach enhances user trust and transparency by making complex recommendation outcomes easier to understand for a lay user.

\end{abstract}

\cortext[cor1]{Corresponding author. Tel.: +0-000-000-0000 ; fax: +0-000-000-0000.}
\end{frontmatter}

\email{author@institute.xxx}




\input{sections/introduction}

\input{sections/relatedwork}

\input{sections/proposedApproach}

\input{sections/evaluation}

\input{sections/risks}
\input{sections/conclusions}

\bibliographystyle{plain}
\bibliography{sn-bibliography}

\end{document}

%% file: sections/introduction.tex
\section{Introduction}\label{sec1}
The introduction of automated decision-making systems in real world contexts reinforces the need to ensure their accountability~\cite{arrieta2020explainable}. Designers of these systems must answer fundamental questions such as: Does the system perform as expected? Do the decisions it makes appear coherent and rational? These concerns have become increasingly important with the widespread adoption of systems based on machine learning~\cite{arrieta2020explainable}. 

In the healthcare field, explainable artificial intelligence (XAI) can be used to interpret and explain decisions made by medical diagnostic models~\cite{litjens2017survey}. For example, a machine learning algorithm designed for skin cancer diagnosis can generate explanations by highlighting specific features or areas of an image that contributed to its classification. These explanations help to understand the reasons that led to a given diagnosis, thus providing the opportunity to validate or reject it, and to make informed decisions based on the results produced by the AI system. The goal of XAI is therefore to design algorithms and models capable of providing not only accurate results, but also understandable explanations of their reasoning~\cite{samek2019explainable}. 
However, the majority of explanations generated by XAI models are primarily intended for experts, highlighting the urgent need to develop more accessible and intuitive forms of explanation, suitable for lay users without advanced technical skills \cite{tchappi2023towards}. This article focuses primarily on the explainability of food recommendation systems for non-expert audience. 

The proposed system leverages the capabilities of Natural Language Processing (NLP), and more specifically on the use of LLMs for generating explanations within recommendation systems. Indeed, the field of NLP has seen remarkable advances in recent years, particularly thanks to the emergence of so-called transformer architectures~\cite{brown2020language}. These architectures rely on attention mechanisms that allow processing to be focused on different parts of the text, thus improving language comprehension and generation. 
From BERT (Bidirectional Encoder Representations from Transformers) to T5 (Text-to-Text Transfer Transformer), GPT (Generative Pre-trained Transformer) to name a few,  progress in LLM is significant~\cite{brown2020language}. 

In this paper, LLM is used to provide personalized food recommendations tailored to each individual's specific dietary needs, health goals, and nutritional requirements~\cite{tessa2023enhancing}. To this end, the proposed system take into account various factors such as age, gender, weight, physical activity level, allergies, and specific food preferences. 
To achieve this goal, our study highlights that the use of LLM, combined with a particular focus on explainability in food recommendation systems, improves the satisfaction and engagement of lay users by providing them with understandable recommendations~\cite{leng2020interpretable}. 
The proposed approach is based on the combination of rule-based models, symbolic reasoning, model visualizations, feature importance analyses, and surrogate models, with the aim of enhancing the explainability of food recommendation systems for a non-expert audience~\cite{sinha2002role}.

This work presents a methodological framework focused on generating natural language explanations of the decisions made by food recommendation systems. The objective is to propose a solution that is both simple and effective, using the prompting technique~\cite{lund2023chatgpt}. We design specific prompts to facilitate the production of concise and informative explanations, highlighting the key features that influenced the system's recommendations. Furthermore, evaluations allow us to compare our approach to existing methods and demonstrate its performance. 

The remainder of this article is structured as follows: Section \ref{sec-explanations} presents the state of the art in XAI applied to recommendation systems. Section \ref{sec:proposedApproach} describes in detail the approach developed in this work, while Section \ref{sec:evaluation} presents its evaluation. Section \ref{sec:risks} highlights the need to strengthen multidisciplinary research around the use of LLM for explaining recommendations, in order to better anticipate and manage potential adverse effects. Finally, Section \ref{sec:conclusion} concludes this paper.

%% file: sections/relatedwork.tex
 \section{Related Works}
\label{sec-explanations}
Recommender systems are widely used in the field of nutrition, with diverse objectives ranging from health promotion to suggesting combinations based on taste preferences. Current approaches to food recommendations generally rely on elements such as recipe content (e.g., ingredients), eating history (e.g., consumption habits), or dietary preferences~\citep{mokdara2018personalized, pagou2023food, sinyabe2023towards}. 
While these approaches are useful, they often lack explanations as to why certain foods or recipes are recommended. This limitation is largely due to the use of "black-box" deep learning models, which do not provide insight into the reasoning behind the proposed recommendations. However, lay users often want to understand the logical or nutritional basis behind the suggestions made.

Recent studies have highlighted the importance of integrating explanations into recommendations to improve their transparency, trust, and acceptability \cite{pagou2023food, tessa2023enhancing}. Although explanations are not yet commonly integrated into food recommendation systems, certain approaches—or combinations of approaches—are attracting increasing interest. 
According to Sovrano and Vitali~~\cite{sovrano2022generating}, the main objective of the generated explanations is to support users in achieving their goals. The authors highlight that users express a higher level of satisfaction when they receive explanations that show them how to achieve their goals. A typical example of this type of explanation would be: ``We recommend this recipe for weight loss because it is low in fat and high in fiber''. This type of justification implicitly answers the question: what should be done to lose weight?, which corresponds to the expectations of a lay user.

In the field of food and culinary arts, several studies have sought to apply logic, reasoning, and query techniques to capture and integrate information from various food-related sources~\citep{haussmann2019foodkg}. 
However, current approaches often rely on predefined explanations that are poorly adapted to users' specific requests \cite{buzcu2022}. Consequently, the possibilities for an individual to request additional clarifications or those better tailored to their personal questions remain limited. 
Furthermore, these explanations are often formulated in a sanitized manner, making them unattractive and uninviting, thus reducing interaction with the XAI model. Under such conditions, the usefulness and effectiveness of the explanatory system are also significantly compromised. 


In light of the above, we propose in this article an approach that differs from previous work by integrating explanations aimed at lay users, relying on hybrid explanatory styles. 
This enriched approach forms the basis for generating detailed explanations, whether to justify the recommendation of a specific food or to answer specific questions related to that recommendation. Indeed, there is a growing recognition of the importance of providing user-centric explanations to build trust and improve understanding of artificial intelligence systems ~\citep{mittelstadt2019explaining}.


%% file: sections/proposedApproach.tex

\section{Proposed Approach}
\label{sec:proposedApproach}

In this Section, we describe the system that combines user-centered and recipe-centered attributes with the capabilities of advanced language models, providing personalized explanations.
 
\subsection{The Baseline Recommendation Strategy}
\subsubsection{Datasets}

Our study leverages two datasets: Recipe Data, extracted from Kaggle's 'foodRecSys-V1', and User Data, gathered through a user data acquisition process. The 'foodRecSys-V1' dataset, containing 49,698 recipes from 'Allrecipes.com', underwent a rigorous cleaning process to yield a final dataset of 17,079 recipes. The User Data, on the other hand, includes anonymized identifiers, demographic information, dietary preferences, and lifestyle details. 

\subsubsection{Food recommender system}

Our research focuses on the development of a knowledge-based Food Recommender System (FRS) that not only offers personalized recipe recommendations but also generates comprehensible explanations for the suggestions made. The principal objective is to provide an answer to the the ``why''  the recommendation was made by the FRS to foster trust and improve lay user experience. 
The heart of our FRS design involves Rule-Based Reasoning (RBR), selected due to its inherent explainability and ability for rule-specific customization. The system gathers user data on dietary preferences, allergies, activity levels, and metabolic energy needs. It then applies explicit rules, devised by nutrition experts, to exclude recipes with restricted ingredients or allergens and calculates daily metabolic energy requirements based on Basal Metabolic Rate (BMR) and activity level to provide tailored meal suggestions.

\begin{figure}[t]
    \centering
\includegraphics[width=.7\linewidth]{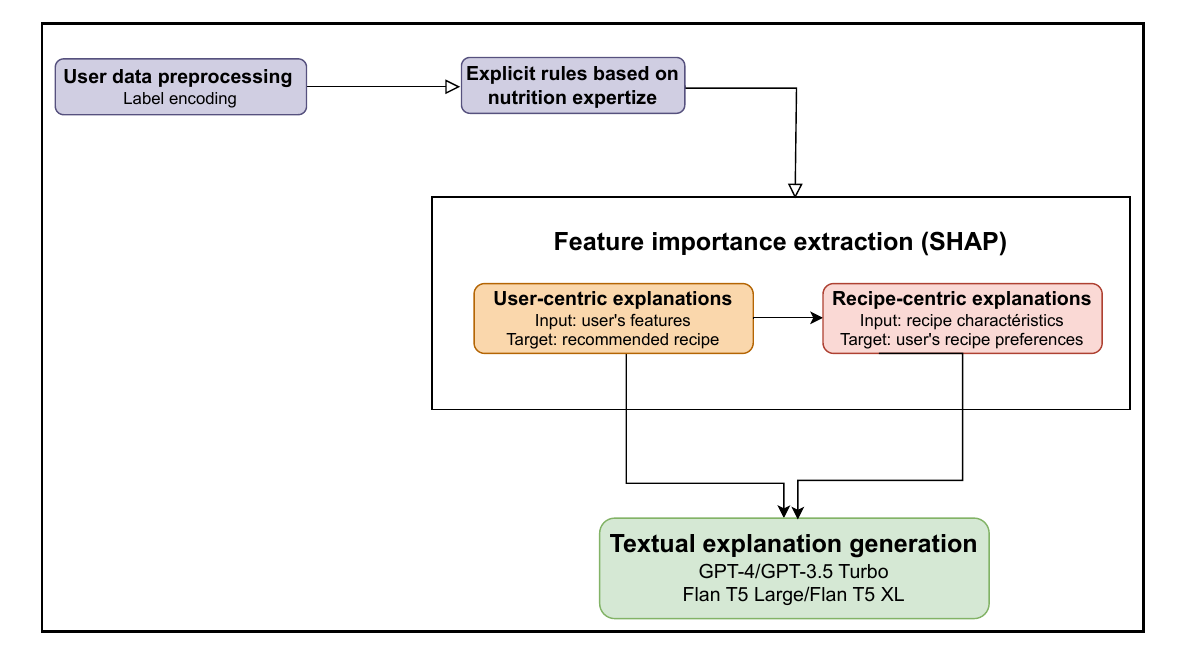} 
 \caption{Overall description scheme.}
    \label{fig:alls}
\end{figure}

Data annotation is pivotal to our FRS, supporting the XAI system. Each user profile is annotated with its corresponding recommended recipe, which enables the XAI system to generate personalized explanations. Additionally, recipes are classified as recommended or not, depending on user preferences, aiding the XAI system to present the reasons behind each recommendation. 
Figure \ref{fig:alls} presents the summary of the approach described in this paper,  composed of the pre-processing steps, the feature importance extraction phase, and the  explanation module. 

\subsection{Post-Hoc Explanation Generation Strategy}


As stated before, our XAI system for meal recommendations is built upon a two-step approach that includes feature importance extraction and textual explanation generation. The first module identifies the most impactful attributes contributing to the final decision of a recipe recommendation. The second module generates varied styles of textual explanations, utilizing the output from the initial module. We introduce innovations such as a post-hoc approach, a combined focus on user and recipe, the use of surrogate models for approximation, and the employment of the SHAP feature importance extraction model. In fact,  we employ a model known as Shapley Additive exPlanations (SHAP) ~\cite{shap} to evaluate the importance of each feature. SHAP uses a game-theoretic approach to compute and assign an ``importance value'' to each feature, indicating how much each contributes to a particular prediction~~\cite{ancona2017towards}. 
By pre-processing user data, we derive the relative importance of user and recipe attributes in influencing the recommendation, offering a detailed and clear understanding of the process behind decision making.

\subsection{Feature Importance Extraction Module} 
In the process of identifying the crucial attributes leading to a specific recommendation within our XAI system, we first undergo a pre-processing stage for the user data. The pre-processing entails transforming string data, such as user demographics, eating habits, and dietary restrictions, into numerical values through a method called ``Label Encoding''. This technique is integral to preparing data for machine learning models, which predominantly function on numerical data.
In the context of our recommendation system, the SHAP values, in correlation with various user attributes and recipe features, can define the influence these characteristics have on a particular recommendation. For example, the time of day or preference for a cuisine type might have higher SHAP values, indicating these factors substantially influence the recommendation.

\subsubsection{User-Centric and Recipe-Centric Features}
In our quest to create an explainable recommendation system, we focus on two types of feature extraction: User-Centric Features and Recipe-Centric Features. 
 Literature indicates that users appreciate explanations that help them understand how to meet their goals. An explanation such as, ``We recommend this recipe to support your weight loss goals because it is low in fats and high in fiber'', serves as an example of a user-centric explanation ~~\cite{wang2021m2lens}. It implicitly addresses the necessary criteria for weight loss, which aligns with the hypothetical goals of the user. 
Conversely, Recipe-Centric Features are targeted towards providing insights into the intrinsic properties of the recommended recipe itself. In the realm of recipe recommendation systems, explanations tend to be user-oriented, justifying why a recipe matches a user's preferences or dietary habits. However, there is a conspicuous lack of exploration into recipe-oriented explanations that provide details about the inherent characteristics of the recipe justifying the recommendation. For instance, ``this recipe is recommended because it has a high protein content and mainly uses seasonal ingredients''.


\subsubsection{Substitution model}
A surrogate model refers to a simplified mathematical construct used for the interpretation of complex models~\cite{audet2000surrogate}. One widely adopted method for explaining black-box models involves the ``single tree approximation''. An example would be approximating a complex  model using a single decision tree.  Our approach uses the first single tree approximation to yield user-centric explanations, and the second one for recipe-centric explanations.  We then synthesize these two explanation types. 

\subsubsection{Use case}
Let's consider the case of our reference lay user, Emna, who is characterized by the following key attributes: she's 30 years old, 170 cm tall, weighs 65 kg, follows a vegetarian diet, has a BMR of 1400 calories, has no food allergies, and expresses a particular aversion to mushrooms. She was recommended a Spaghetti Carbonara recipe. 

To elucidate the reasoning behind this recommendation, we employ two distinct surrogate models, each represented by a decision tree. The first decision tree takes Emna's features as input and the recommended recipe as a target, using the SHAP method to determine the relative importance of user-centered characteristics. For instance, Emna's vegetarian diet might have a relative importance of 0.5, her BMR 0.2, her dislike for mushrooms 0.3, and so on.
Then, a second decision tree is used, which takes recipe characteristics as input, with Emna's recipe preferences as targets. Again, we use SHAP to ascertain the importance of recipe-centered characteristics.
By juxtaposing these two sets of feature importance, we aim to provide a detailed yet easily comprehensible explanation of why a particular recipe was recommended to Emna. For instance, we could say, `` This recipe was recommended because it aligns with your vegetarian diet, does not contain mushrooms which you dislike, and has a high protein and low sugar content. It is also rich in fiber, quick to prepare, and highly rated''. This provides a comprehensive overview of the recommendation process.

\subsection{Textual Explanation Generation Module}

Our Recommendation Explanation Module operates on a unique blend of user- and recipe-focused attributes to generate robust, personalized explanations. It is empowered by advanced language models, including GPT-4, GPT-3.5 Turbo, Flan T5 Large, and Flan T5 XL.
For the user-focused attributes, the module considers elements such as health goals, dietary restrictions, or specific dislikes, ensuring the recommendation aligns with the user's unique lifestyle and preferences. 
On the other hand, recipe-focused attributes might involve nutritional details, preparation time, or specific ingredients, highlighting why the suggested recipe is the best choice considering its intrinsic qualities~\cite{longpre2023flan} .

These attributes are integrated within an  engineered prompt: ``Convince me that 'recipe name' is better for me, given $attribute_1^{recipe}$: $value_1$,  
$attribute_2^{recipe}$ : $value_2$, ..., $attribute_n^{recipe}$ : $value_n$''. This prompt leverages the persuasive power of our LLMs to generate compelling justifications for the recommended recipe.
The explanations generated are not only detailed but also understandable by lay users, underscoring why the recommended recipe suits the user, considering both their personal preferences and the inherent qualities of the recipe. 


%% file: sections/evaluation.tex
\section{Evaluation}
\label{sec:evaluation}

To evaluate the acceptability and effectiveness of our proposed explanation generation framework, we conducted a survey to gather human evaluations of the explanations. 
In the survey, the participants (a total of 19) were asked to rate the explanations provided for each recommended recipe on a Likert 5-scale. Additionally, participants were given the opportunity to choose their preferred explanation and provide a detailed explanation for their choice. To this end, we generated plain and contrastive explanations using four different models: FlANt5large (Model 1), FlanT5XL (Model 2), GPT3.5 (Model 3), and GPT4 (Model 4).
To analyze the result of the conduct survey through three major axes : (i) We compute and compare the average ratings of each model's generations for both plain and contrastive explanations seperately and together, (ii) We compute and analyse the pourcentage of prefered models' generations for both plain and contrastive explanations seperately and together, (iii) We analyse the reasons behind the best model choice given by the users to get better insight into the effectiveness of each model.

\subsection{Average rating of models explanations}

\begin{figure}[h!]
    \centering
    \includegraphics[width=0.35\textwidth]{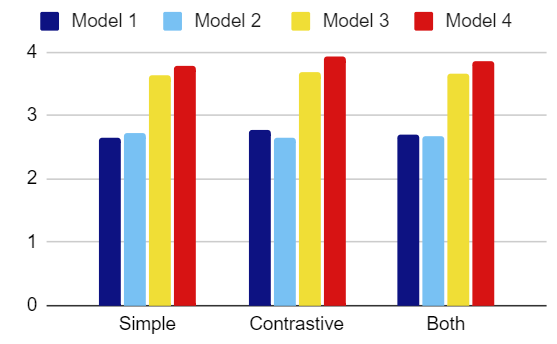}
    \caption{Average rating of models simple, contrastive and both explanations.}
    \label{fig:all}
\end{figure}

After collecting ratings from all participants regarding each model's plain and contrastive explanations for each user recipe recommendation, we calculated the average ratings and presented the results in the histogram shown in Figure \ref{fig:all}. It displays the average ratings for each model's performance, both with plain and contrastive explanations, separately and combined.

The analysis reveals that all models received an average rating greater than 2.5 on a 5-point scale, confirming the validity of our approach. Notably, the GPT models stood out with an average rating approaching 4 out of 5, showcasing the high effectiveness of this approach when utilizing GPT models.
A more detailed examination shows that the GPT models outperformed the FLAN models by almost one point. Specifically, GPT 3.5 and GPT 4 demonstrated similar performance, with GPT 4 slightly edging ahead in all cases. This finding highlights the incremental improvement achieved in GPT versions. 
Overall, these results affirm the strength of our chosen approach and underscore the promising capabilities of GPT models in this context. 

\begin{figure}[h!]
    \centering
    \begin{minipage}{0.4\linewidth}
        \centering
        \includegraphics[width=.8\linewidth]{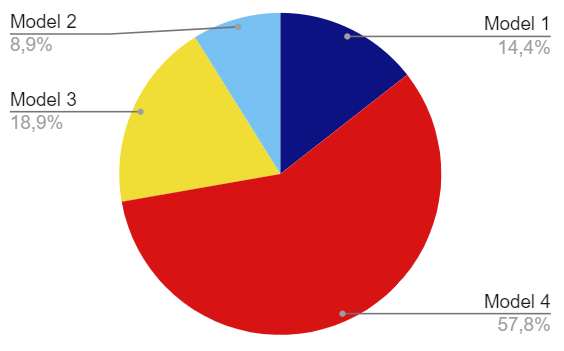}
        \caption{Pourcentage of preferred models' contrastive explanations.}
        \label{fig:contrastive}
    \end{minipage}\hfill
    \begin{minipage}{0.4\linewidth}
        \centering
        \includegraphics[width=.8\linewidth]{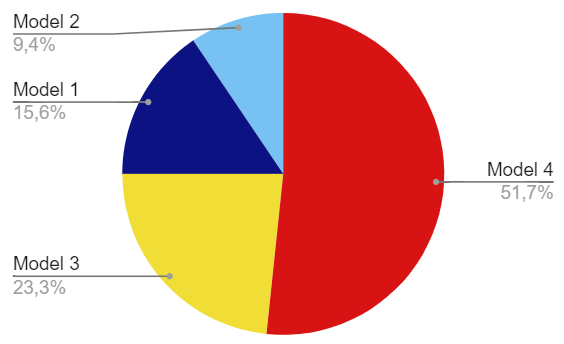}
        \caption{Pourcentage of preferred models' plain explanations.}
        \label{fig:plain}
    \end{minipage}
\end{figure}


Furthermore, we computed the percentage of each model appearance. Figure \ref{fig:plain} presents the distribution of preferred models for plain explanations. Interestingly, the participants' top choice was the GPT 4 model, accounting for an impressive 51.7 \% selection percentage. Following closely behind were the GPT 3.5 model withh 23\%, which garnered significant support, and the Flan T5 models, which received less consideration.
The analysis further extended to contrastive explanations, as showcased in Figure \ref{fig:contrastive}. Here, the GPT 4 model exhibited the best performance across all cases. This finding indicates a consistent preference for GPT4's contrastive explanations among the participants. 
These results shed light on the added value of the GPT 4 model for generating explanations to lay users, both in terms of plain and contrastive explanations, as the most preferred choice among users. The high selection percentage for GPT 4 underscores its effectiveness and suggests that it might be the most suitable model for generating explanations in this context. 


\subsubsection{Further Analyzing the Participants remarks}

The participants compared the explanations provided by each model and expressed their preferences based on various criteria. In fact, 5 participants were selected for further interviews to understand their choices. Table \ref{tab:whys}, presents some of the participants reasons of choosing a specific model.






\begin{table}[h]
\caption{Participants remarks}
\scriptsize
  \label{tab:whys}
  \centering
  \begin{tabularx}{\linewidth}{p{1.5cm}|X}
    \hline
     Preferred Model & Reason \\
    \hline
    Model 1 & The first model gave us the info we need, whereas second model didn't give much info. third and last models  elaborate more than it's required. \\
    
    Model 4 & In contrast to the first two models, model 4 offers a more human-like description, employing conversational language and skillfully integrating persuasive words with greater precision and brevity compared to the third model. \\
    
    Model 3 & While most models gave the essential information about the recipe, model 3 sold it better. \\
    
    Model 2 & It gave the more useful explanations in my opinion. \\
    
    Model 3 & Again I prefered both the third and fourth models' answers and chose the third one because the fourth one sounded more like a commercial. 
    \\   
    \hline
\end{tabularx}
\end{table}


From these interviews we can learn that: (i) Participants liked personalized explanations that include relevant statistics and  individual preferences, (ii) A balance between concise and comprehensive explanations is needed.

%% file: sections/risks.tex
\section{Potential Second-order Risks of the Use of LLMs in XAI}
\label{sec:risks}

The opportunities offered by the use of LLMs in XAI models have been enumerated. In particular, it has been highlighted how they can benefit the user by enhancing the effectiveness and efficiency of explanations. Nevertheless, this is not enough to ignore the limitations that still accompany LLMs, as well as the possibility that they expose the individuals involved to negative effects, even if not expected or desired by programmers and designers. In particular, second-order effects are defined as any effect that is not the intended or specifically expected one, but which is an indirect or secondary consequence of the effort put into realizing the primary effects~\citep{sunstein1999second}. 
In the context of AI, second-order effects are more commonly referred to as risks, even unplanned outcomes, which may nevertheless derive from the design and programming expedients with which AI systems are implemented. 

With respect to the use of LLMs in XAI, more specifically, the main risks for the end user are represented by the discrepancy that is often created between the perceived capabilities of the LLM and the actual ones. 

A primary illustration concerns the capacity of LLMs to adjust their conversational style in accordance with the subject's inquiries, while aligning with the context, personality, and general aptitude of the individual involved in the interaction. This frequently results in the supposition that the system possesses the capability to develop a comprehensive theory of the user's cognitive processes and personality~\cite{castano2023listening}. However, extant studies have largely failed to support this perception~\cite{strachan2023testing}. This misconception of LLMs' capabilities could more easily lead users into a state of overtrust. Given the system's apparent ability to interact appropriately and engagingly with the human interlocutor, sometimes even mentioning concepts, experiences, or expressions familiar to the person themselves, could lead the latter to confer a level of reliability and infallibility to the system which is both higher than the real and desirable one~\cite{kim2024m}. In the context of XAI, it is imperative for users to maintain criticality in evaluating the explanations provided, assessing their persuasiveness and accuracy.  Failure to do so can result in explanations that lack transparency, merely serving to justify the recommendations made and subconsciously influencing acceptance without full awareness. Indeed, studies have demonstrated that design techniques --- especially when in connection with the use of familiar, engaging natural language, and a marked interactivity of the systems --- can induce forms of subliminal persuasion that risk resulting in manipulation of the user~\cite{carli2022human}. 

Furthermore, is important to note that LLMs frequently adapt to the initial or ingrained beliefs exhibited by the user~\cite{strachan2023testing}, rather than introducing new information or data for the autonomous evaluation of individuals. This tendency is especially pronounced when the user challenges the answer or, as in the case analysis, the explanation provided by the system. When users are firmly entrenched in their beliefs, LLMs are inclined to modify their responses to align with the user's viewpoint, even if it is erroneous and can be disproven based on the data the LLM should posses. 

The aforementioned points highlight the crucial importance of multidisciplinary research in XAI, particularly in the context of LLMs' utilization within XAI. Through the joint effort of 
experts from diverse fields, including neuroscience, cognitive science, ethics, and law, who engage in collaborative teams with designers and developers, it becomes more feasible to illuminate the challenges associated with LLMs' deployment in highly interactive systems. Thus, it could be also possible to weigh up design and programming choices so as to reduce the occurrence of negative second-order effects, or to develop counterbalance methods.

%% file: sections/conclusions.tex
\section{Conclusions}\label{sec:conclusion}

The increasing emphasis on accountability within the sphere of automated decision-making systems led to XAI. 
The framework presented in this paper introduces a  solution leveraging LLMs to the generation of explanations of a recipe recommended in natural language to lay users. The explanations aims to foster user trust and enables informed dietary choices, culminating in heightened satisfaction, engagement, and holistic wellness. 


Future research can further fine-tune and broaden the proposed framework, incorporating other techniques and applications to bolster the elucidation and user experience of decision-making systems.
In addition, multidisciplinary research in this field needs to be strengthened in order to ensure that LLM used to enhance the positive effects of XAI do not hinder users with negative secondary effects, which are still widely underestimated.